\documentclass[prl,twocolumn,nofootinbib,reprint]{revtex4-1}
\usepackage{amsmath,amssymb,amsthm,bm,braket,ascmac,bbm}
\usepackage{graphicx}
\usepackage{color} 
\usepackage{hyperref}
\usepackage{amssymb}
\usepackage{tabularx}

\theoremstyle{definition}

\begin{document}
\hfill KEK-TH-2270

\title{
Implications of Gravitational Waves for Supersymmetric Grand Unification
}
\author{ So Chigusa$^{1,2,3}$, Yuichiro Nakai$^4$, and Jiaming Zheng$^4$}
\affiliation{\vspace{2mm} \\
$^1$Berkeley Center for Theoretical Physics, Department of Physics,\\
University of California, Berkeley, CA 94720, USA \\
$^2$Theoretical Physics Group, Lawrence Berkeley National Laboratory, Berkeley, CA 94720, USA \\
$^3$KEK Theory Center, IPNS, KEK, Tsukuba, Ibaraki 305-0801, Japan \\
$^4$Tsung-Dao Lee Institute and School of Physics and Astronomy, \\
Shanghai Jiao Tong University, 800 Dongchuan Road, Shanghai, 200240 China 
}

\begin{abstract}

Supersymmetric grand unification based on $SO(10)$ is one of the most attractive paradigms
in physics beyond the Standard Model. 
Inspired by the recent NANOGrav signal,
we discuss the implications of detecting a stochastic gravitational wave background
emitted by a network of cosmic strings for the $SO(10)$ grand unification.
Starting from a minimal model with multiple steps of symmetry breaking, we show that it generally prefers a high intermediate scale above $10^{14}$\,GeV that is favored by observable primordial gravitational waves.
The observed spectrum can potentially narrow the possible range of the cosmic string scale and restricts the unified couplings and the unification scale by requiring gauge coupling unification. As an indirect consequence of the high cosmic string scale, the monopole abundance places non-trivial constraints on the theory. These are complementary to the proton decay constraints and probe different facets of supersymmetric SO(10) unification theories.  


\end{abstract}

\maketitle

{\bf Introduction.--}
The structure of the Standard Model (SM) matter sector, quarks and leptons, and
the high-energy behavior of the SM gauge couplings strongly suggest that
the three SM gauge groups $G_{\rm SM} \equiv SU(3)_C \times SU(2)_L \times U(1)_Y$
are unified at a high-energy scale
\cite{Georgi:1974sy}.
The smallness of neutrino masses further indicates the existence of heavy right-handed neutrinos
to realize the seesaw mechanism \cite{Yanagida:1980xy,Minkowski:1977sc,GellMann:1980vs},
which is a consequence of the grand unification based on $SO(10)$
\cite{Fritzsch:1974nn,Georgi:1974my}.
A large hierarchy of scales between the electroweak symmetry breaking (EWSB)
and the grand unification is naturally stabilized by supersymmetry (SUSY).
Amazingly, in the minimal supersymmetric SM (MSSM),
the precise unification of the three SM gauge couplings is achieved.
Moreover, with appropriate choices of Higgs representations to break the $SO(10)$,
a $\mathbb{Z}_2$ subgroup known as the matter parity remains unbroken at all scales \cite{Krauss:1988zc,Ibanez:1991hv,Ibanez:1991pr,Martin:1992mq},
while such parity is an ad-hoc global symmetry in the MSSM.
The matter parity forbids the rapid decay of protons
and ensures the stability of the lightest supersymmetric particle which gives a viable candidate
of the dark matter\,\cite{Goldberg:1983nd,Ellis:1983ew}.
Therefore, the supersymmetric $SO(10)$ grand unified theory (GUT) is one of the most compelling frameworks of
physics beyond the SM.


Since the grand unification is realized at a very high energy scale,
its experimental test must be indirect.
The most famous prediction of the GUT is the finite lifetime of the proton.
The current lower limit from the Super-Kamiokande experiment is
$1.6 \times 10^{34} \, \rm years$ \cite{Miura:2016krn} for the $p\to\pi^0 e^{+}$ decay mode
and $5.9\times 10^{33}\, \mathrm{years}$ \cite{Abe:2014mwa} for the $p\to K^{+}\bar{\nu}$ mode.
In the near future, the limit is expected to be improved to $7.8 \times 10^{34} \, \rm years$ ($3.2\times 10^{34}\,\mathrm{years}$) for the $\pi^0 e^{+}$ ($K^{+}\bar{\nu}$) mode
at the Hyper-Kamiokande experiment \cite{Abe:2018uyc}.
Another hint may come from topological defects associated with GUT phase transitions.
In particular, a network of cosmic strings can be produced and the matter parity renders them stable\,\cite{Kibble:1976sj,Kibble:1982ae,Jeannerot:2003qv}.
The cosmic strings form closed loops, shrink and lose energy via the emission of gravitational waves (GWs) \cite{Vilenkin:1981bx,Vachaspati:1984gt}.\footnote{Numerical simulations based on Nambu-Goto strings
support this picture
\cite{Ringeval:2005kr,BlancoPillado:2011dq}.
In the Abelian Higgs model, however, cosmic strings lose energy via particle productions,
which suppress the GW production
\cite{Hindmarsh:2011qj,Buchmuller:2013lra}.
Recent simulations indicate that such particle productions are important only for very small loops
\cite{Matsunami:2019fss}.}
Interestingly, a stochastic GW background produced by cosmic strings
stretches across a wide range of frequencies,
and such a GW signal is one of the main targets in multi-frequency GW astronomy and cosmology. 

Recently, the NANOGrav collaboration has reported the first evidence of a stochastic GW background
in pulsar timing data
\cite{Arzoumanian:2020vkk}.
Although the signal is not conclusive,
it can be well fitted by GWs from a network of cosmic strings because they can give a favored flat spectrum of frequencies in the GW energy density
\cite{Ellis:2020ena,Blasi:2020mfx,Buchmuller:2020lbh}.\footnote{Other interpretations of sources to generate GWs to explain the NANOGrav signal
include phase transitions
\cite{Nakai:2020oit,Addazi:2020zcj,Ratzinger:2020koh,Neronov:2020qrl,Bian:2020bps},
primordial black hole formation \cite{Vaskonen:2020lbd,DeLuca:2020agl,Kohri:2020qqd,Sugiyama:2020roc,Domenech:2020ers,Inomata:2020xad}
and dynamics of axion-like particles \cite{Ratzinger:2020koh,Namba:2020kij}.}
The energy density spectrum is proportional to $(G\mu)^2$ where
$G$ is the Newton's constant and $\mu$ is the cosmic-string tension.
The spectrum also depends on the loop size at the time of formation $\alpha$.
This parameter has not been reliably estimated so far,
but larger loop size $\alpha = 0.01-0.1$ is typically favored by recent numerical simulations\,\cite{Ellis:2020ena,Blanco-Pillado:2013qja,Blanco-Pillado:2017oxo}. 
Conservatively, taking $\alpha = 3 \times 10^{-4}, 5 \times 10^{-3}, 1 \times 10^{-1}$,
the NANOGrav signal can be fitted with
$G\mu = 1 \times 10^{-7} , 5 \times 10^{-9}, 1 \times 10^{-10}$, respectively
\cite{Blasi:2020mfx}.
The symmetry breaking scale associated with the formation of cosmic strings
is related to the string tension,
$v \sim 10^{16} \, {\rm GeV} \left( {G\mu}/{10^{-7}} \right)^{1/2}$,
whose precise coefficient is given by an $\mathcal{O}(1)$ group theory factor.
Then, the NANOGrav signal indicates the symmetry breaking scale is in the range,
$10^{14} \, {\rm GeV} \lesssim v \lesssim 10^{16} \, {\rm GeV}$.


The interpretation of the NANOGrav signal with cosmic strings from a GUT phase transition indicates the existence of intermediate steps\,\cite{Chakrabortty:2020otp} in the breaking of $SO(10) \rightarrow G_{\rm SM}\times M$ (with $M$ the matter parity)
because the $SO(10)$ breaking also predicts monopoles that may overclose the universe.
In a natural cosmologically safe scenario, monopoles are only produced during symmetry breaking at high energy scales and
get diluted away by inflation afterward. The remaining intermediate gauge group is finally broken to $G_{\rm SM}$
at a lower scale $v\gtrsim 10^{14}$\,GeV where cosmic strings emitting GWs are formed.

In this paper, we study the consequences of such intermediate scales in a supersymmetric $SO(10)$ theory.
The particles with masses below the unification scale largely alter the renormalization group (RG) evolution of the gauge couplings from that in MSSM.
Interestingly, it turns out that the unification of gauge couplings enforces high intermediate scales that are needed by the cosmic string interpretation of the NANOGrav result.
Thus, the observation of cosmic string GW does not only imply the existence of an intermediate scale but also strongly motivates supersymmetric SO(10) as the unified theory of gauge interactions.


{\bf Models.--}
$SO(10)$ is a rank 5 simple group and contains an additional $U(1)$ factor in addition to $G_{\rm SM}$ of rank 4.
The minimal choices of Higgs representations to break the $U(1)$ in a SUSY model are ${\bf 16}+{\bf \overline{16}}$ or ${\bf 126}+{\bf \overline{126}}$. We will stick to the latter choice as it preserves the matter parity $M\equiv (-1)^{3(B-L)}$ where $B$ and $L$ are baryon and lepton numbers.
Vacuum expectation values (VEVs) of SM singlets in ${\bf 126}+{\bf \overline{126}}$ leave
the $SU(5)$ subgroup unbroken and more Higgs fields are needed.
The minimal Higgs fields that break $SO(10)$ to $G_{\rm SM}\times M$ are then ${\bf 45}+{\bf 54}+{\bf 126}+{\bf \overline{126}}$\,\footnote{The choice of ${\bf 45}+{\bf 54}+{\bf 126} +{\bf \overline{126}}$ is ``minimal" in the number of components. Another popular choice is ${\bf 210}+{\bf 126}+{\bf \overline{126}}$ \,\cite{Bajc:2004xe,Aulakh:2003kg} which leads to the ``minimal" number of parameters in the general Lagrangian.} and we will focus on this choice throughout the paper.
The two Higgs doublets $H_u, H_d$ in the MSSM are chosen as linear combinations of bi-doublets in $\bf 10$ and $\bf 120$ to generate the correct SM Yukawa couplings and neutrino masses.\footnote{
Without $\bf 120$, the $\bf 10$ and $\bf \overline{126}$
Yukawa couplings are constrained by the SM fermion masses.
The resultant right-handed neutrinos are too heavy
and contradict the neutrino oscillation measurements\,\cite{Melfo:2010gf}. For minimal fine-tuning conditions, we assume that components other than $H_u, H_d$ in $\bf 10$ and $\bf 120$ stay at the GUT scale.}

To summarize, we consider SUSY $SO(10)$ models with the following set of heavy Higgs fields
\cite{Aulakh:2000sn},
\begin{align}
    S=\mathbf{54},~~A=\mathbf{45},~~
    \Sigma=\mathbf{126},~~ \overline{\Sigma}=\overline{\mathbf{126}}.
\end{align}
The most general renormalizable superpotential of the Higgs sector is
\begin{align}
    W_H &= \frac{m_S}{2} \mathrm{Tr}\,S^2
    + \frac{\lambda_S}{3} \mathrm{Tr}\, S^3
    + \frac{m_A}{2} \mathrm{Tr}\, A^2
    + \lambda \mathrm{Tr}\, A^2 S
    \notag \\
    &+ m_\Sigma \Sigma \overline{\Sigma} 
    + \eta_S \Sigma^2 S
    + \overline{\eta}_S \overline{\Sigma}^2 S
    + \eta_A \Sigma \Bar{\Sigma} A
    + \cdots\, ,
\label{eq:SO10_lagrangian}
\end{align}
where we have omitted the terms with $\bf 10$ and $\bf 120$.
We find minima of the Higgs potential by solving the $F$- and $D$-flat conditions.
We express the VEVs of $G_{\mathrm{SM}}$ singlet fields
in terms of representations under the Pati-Salam group $G_{422}\equiv SU(4)_C\times SU(2)_L \times SU(2)_R$ to which they belong:
\begin{align}
    s &= \braket{S(1,1,1)},~
    a = \braket{A(15,1,1)},~
    b = \braket{A(1,1,3)},
    \notag \\
    \sigma &= \braket{\Sigma(10,1,3)},~
    \overline{\sigma} = \braket{\overline{\Sigma}(\overline{10},1,3)}.
    \label{eq:vevs}
\end{align}
Patterns of VEVs which satisfy the minimum condition and lead to a subgroup $H$ of $SO(10)$ are summarized in Table~\ref{tab:vacuum}.
Note that the $D$-flatness sets $|\sigma| = |\overline{\sigma}|$.


\begin{table}[t]
\renewcommand{\arraystretch}{1.5}
\newcolumntype{C}{>{\centering\arraybackslash}X}
    \begin{tabularx}{80mm}{c|cXXXX}
        $H$ &&  $s$  & $a$ & $b$ & $\sigma$ \\
        \hline
        $SO(10)$ &&  &  &  &  \\
        $\circ$ $SU(4) \times SU(2) \times SU(2)$ && $\checkmark$ &&  & \\
        $\star$ $SU(3) \times SU(2) \times SU(2) \times U(1)$ && $\checkmark$ & $\checkmark$ &  &  \\
        $\star$ $SU(4) \times SU(2) \times U(1)$ && $\checkmark$ &  & $\checkmark$ &  \\
        $SU(5) \times M$ && & $\checkmark$ & $\checkmark$ & $\checkmark$ \\
        $G_{\rm SM} \times M$ && $\checkmark$ & $\checkmark$ & $\checkmark$ & $\checkmark$ \\
    \end{tabularx}
    \caption{
        The summary of patterns of VEVs at local minima of the potential that lead to a subgroup $H$ of $SO(10)$ ($\checkmark$ denotes a nonzero VEV).
        Stars in the first column indicate that the symmetry breaking $H \to G_{\mathrm{SM}} \times M$ is associated with the formation of cosmic strings.
        A circle in the first column indicates the formation of monopoles and cosmic strings at the same time, which is not of our interest.}
    \label{tab:vacuum}
\end{table}

\begin{table*}[t]
\renewcommand{\arraystretch}{1.4}
    \centering
    \begin{tabular}{c|c}
        States & Mass scale \\
        \hline
        $\Sigma(\overline{10},3,1)$, $\overline{\Sigma}(10,3,1)$ & $M_C$ \\
        color triplets and sextets of $\Sigma(10,1,3)$, $\overline{\Sigma}(\overline{10},1,3)$ & $M_C$ \\
        color triplets of $A(15,1,1)$ & $M_C$ \\
        $(1,1)_0$, $(1,1)_{\pm 1}$ from $\Sigma(10,1,3)$, $\overline{\Sigma}(\overline{10},1,3)$ & $M_R$\\
        a color octet and a singlet of $A(15,1,1)$ & $M_1 \equiv \mathrm{Max}\left[ \frac{M_R^2}{M_C}, \frac{M_C^2}{M_X} \right]$ \\
        $(1,1)_0$, $(1,1)_{\pm 2}$ from $\Sigma(10,1,3)$, $\overline{\Sigma}(\overline{10},1,3)$ & $M_2 \equiv M_R^2/M_X$\\
        all the other components & $M_X$ \\
    \end{tabular}
    \caption{The mass spectrum of $S$, $A$, $\Sigma$, and $\overline{\Sigma}$ Higgs fields in the model A.
    $(1,1)_{0,\pm 1, \pm 2}$ denotes charges under the SM gauge groups
    $G_{\rm SM} \equiv SU(3)_C \times SU(2)_L \times U(1)_Y$.}
    \label{tab:mass-spectrum}
\end{table*}

\begin{table}[t]
    \renewcommand{\arraystretch}{1.4}
    \centering
    \begin{tabular}{c|c|c|c|c}
        Energy range & $b_1^{(k)}$ & $b_2^{(k)}$ & $b_3^{(k)}$ 
        & $(k)$ 
        \\ \hline 
        $M_Z < Q < M_S $ & $41/10$ & $-19/6$ & $-7$ & $(1)$ \\
        $M_S < Q < M_2 $ & $33/5$ & $1$ & $-3$ & $(2)$ \\
        $M_2 < Q < M_C $ & $57/5-12/5\theta_R$ & $1$ & $-3+3\theta_1$ & $(3),(4),(5)$ \\
        $M_C < Q < M_X $ & $191/5$ & $41$ & $34$ & $(6)$
    \end{tabular}
    \caption{The coefficients of the RG equations of
    the gauge coupling constants for each energy range in the model A.
    Note that $\theta_R \equiv \Theta(Q-M_R)$ and $\theta_1\equiv \Theta(Q-M_1)$ with $\Theta$ being the Heaviside step function.}
    \label{tab:RGE}
\end{table}

Motivated by the result of NANOGrav, we focus on symmetry breaking patterns where cosmic strings form at an intermediate scale $M_R \sim 10^{\operatorname{14-16}}\,\mathrm{GeV}$. 
As indicated by stars in the first column of Table~\ref{tab:vacuum}, there are two possible choices of $H$ whose breaking into $G_{\mathrm{SM}} \times M$ is associated with the cosmic string formation without accompanying monopoles: 
$G_{3221}\equiv SU(3)_C \times SU(2)_L \times SU(2)_R \times U(1)_{B-L}$ and $G_{421}\equiv SU(4)_C \times SU(2)_L \times U(1)_R$.
When there is a hierarchy between VEVs of Eq.~\eqref{eq:vevs}, the multi-step breaking of $SO(10)$ with at least one intermediate scale is achieved.
The simplest possibility is the breaking with one intermediate scale,
i.e., $SO(10) \to H \to G_{\mathrm{SM}} \times M$, where the breaking scale of $H$ is identified as $M_R$.
A detailed analysis of this possibility is given in the appendix, where we show that
it is difficult to construct a viable model that survives collider constraints of the SUSY particle search.
Thus, in the following discussion, we focus on models with two intermediate scales, with the lower scale identified as $M_R$, and show that $M_R \gtrsim 10^{14}\,\mathrm{GeV}$ is compatible with the unification of gauge couplings.
There are only two possible breaking patterns that lead to surviving cosmic strings, $SO(10)\to G_{422}\to H\to G_{\mathrm{SM}}\times M$ with $H=G_{3221}$ (model A) or $H=G_{421}$ (model B).

The model A is obtained through the hierarchical choice of VEVs given by $|s| \gg |a| \gg |\sigma|$, while $|b|\sim |\sigma|^2/|s| \ll |\sigma|$ is ensured from the minimization condition of the potential.
We can define the unification scale $M_X$, the Pati-Salam breaking scale $M_C$, and the
$G_{3221} \rightarrow G_{\mathrm{SM}}\times M$ breaking scale $M_R$ through $M_X \sim |s|$, $M_C \sim |a|$, and $M_R \sim |\sigma|$, respectively.
We summarize the mass spectrum of all the components of $S$, $A$, $\Sigma$, and $\overline{\Sigma}$ Higgs fields in Table~\ref{tab:mass-spectrum}.\footnote{
Note that there is mixing among the Higgs doublets in $\bm{10}$, $\Sigma$, and $\overline{\Sigma}$, and two of them obtain light masses to be MSSM Higgs doublets.
The other Higgs doublets obtain masses of $\mathcal{O}(M_X)$, which are summarized in the last column of Table.~\ref{tab:mass-spectrum}
.
}
Note that some states have masses different from the scales $M_{X, C, R}$.
The mass scale $M_2$ is always smaller than $M_{1, R}$, while the order of $M_1$ and $M_R$ is not fixed in general.
When we perform multiple steps of RGE running, we assume all particles at an energy scale $Q$ share the same mass $Q$. Generally, their masses depend on coupling constants and the non-uniform spectrum generates a threshold correction
\cite{Weinberg:1980wa,Hall:1980kf}.
We discuss this point in appendix.


The model B is obtained when $|s| \gg |b| \gg |\sigma|$ and $|a|\sim |\sigma|^2/|s| \ll |\sigma|$.
However, it turns out that typically the unification scale is lower in this model than that in the model A.
As a result, the current constraint from the proton decay rate is severer and only a tiny region remains unconstrained.
Thus, we will not consider this possibility and focus on the model A in the following discussion.

{\bf Gauge coupling unification.--}
Let us now consider the RG evolution of the gauge couplings from the unification scale $M_X$ to the weak scale $M_Z$ experiencing $SO(10)\to G_{422}\to G_{3221} \to G_{\mathrm{SM}}\times M$.
We assume the mass scale of supersymmetric particles $M_S$ is smaller than $M_2$.
Then, the evolution of the three SM gauge couplings $\alpha_i \equiv g_i^2 / 4\pi$ $(i=1,2,3)$ is governed by the SM and MSSM beta functions up to $M_S$ and $M_2$, respectively, while light states from GUT Higgses contribute to the running from $M_2$ to $M_R$.
At the $G_{3221}$ breaking scale $M_R$,
the gauge couplings of $SU(2)_R$ and $U(1)_{B-L}$, denoted as $\alpha_{2R}$ and $\alpha_{B-L}$, respectively, are matched to that of the SM hypercharge with%
\footnote{The convention for the charge of $U(1)_{B-L}$ is\, $q_{B-L}=\frac{3}{8}(B-L)$.}
\begin{align}
    \frac{3}{5} \alpha_{2R}^{-1} (M_R) + \frac{2}{5} \alpha_{B-L}^{-1} (M_R) = \alpha_1^{-1} (M_R),
    \label{eq:matching-3221}
\end{align}
and $\alpha_{2R}$ and $\alpha_{B-L}$ run further up to the $G_{422}$ breaking scale $M_C$.
The matching condition at $M_C$ is
\begin{align}
    \alpha_4 (M_C) = \alpha_3 (M_C) = \alpha_{B-L} (M_C),
\end{align}
where $ \alpha_4$ denotes the $SU(4)_C$ gauge coupling.
Finally, $\alpha_4$, $\alpha_2$, and $\alpha_{2R}$ run to $M_X$ and unify into a unique value $\alpha_U$.
At the one-loop level,
the relationship between the gauge coupling constants $\alpha_i$ at $M_Z$ and $\alpha_U$
is described by
\begin{align}
    \frac{2\pi}{\alpha_i (M_Z)} &= \frac{2\pi}{\alpha_U} + \left[
    b_i^{(1)} \ln\frac{M_S}{M_Z}
    + b_i^{(2)} \ln\frac{M_2}{M_S}
    + b_i^{(3)} \ln\frac{M_1}{M_2} \right. \notag \\
    &
    \left.+ b_i^{(4)} \ln\frac{M_R}{M_1}
    + b_i^{(5)} \ln\frac{M_C}{M_R}
    + b_i^{(6)} \ln\frac{M_X}{M_C}
    \right],
\end{align}
for $M_1 < M_R$, while the role of $M_1$ and $M_R$ should be interchanged if $M_1 > M_R$.
The coefficients $b_i^{(k)}$ are summarized in Table~\ref{tab:RGE}. In particular, $b_i^{(3,4,5)}$ are written in a compact form within $M_2 < Q < M_C $ with thresholds at $M_1$ and $M_R$ represented by step functions $\theta_1$ and $\theta_R$.
We solve these three equations in terms of the three parameters
$M_{X,C,R}$ and obtain a set of solutions as functions of $M_S$ and $\alpha_U$.
In our numerical analysis, we use the two-loop RG equations\,\cite{Arason:1991ic,Martin:1993zk} below $M_R$ while including only the one-loop contributions from the light states.

\begin{figure*}[t]
\centering
\begin{minipage}[t]{0.32\hsize}
    \centering
    \includegraphics[width=1\linewidth]{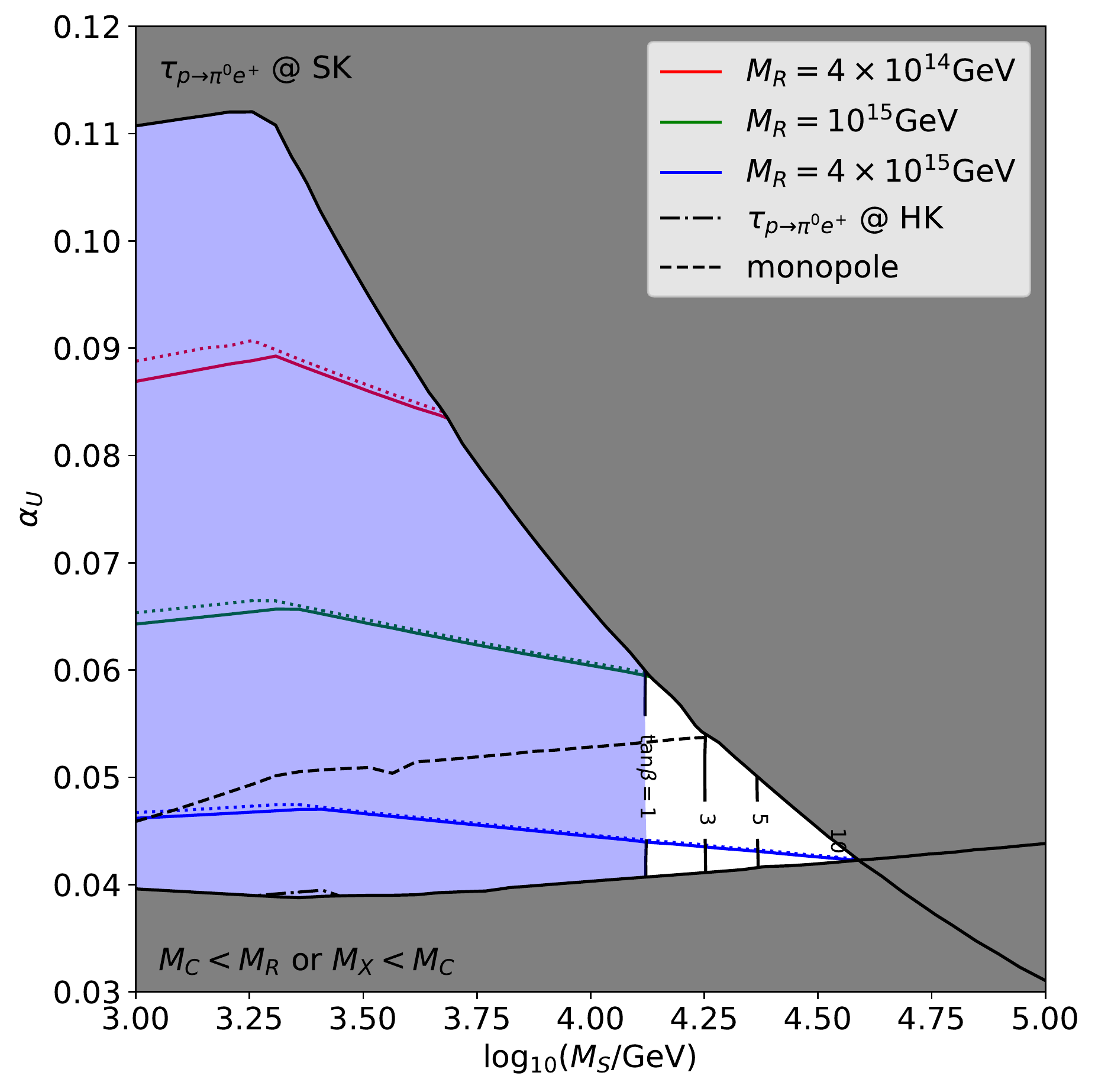}
\end{minipage}
\begin{minipage}[t]{0.32\hsize}
    \centering
    \includegraphics[width=1\linewidth]{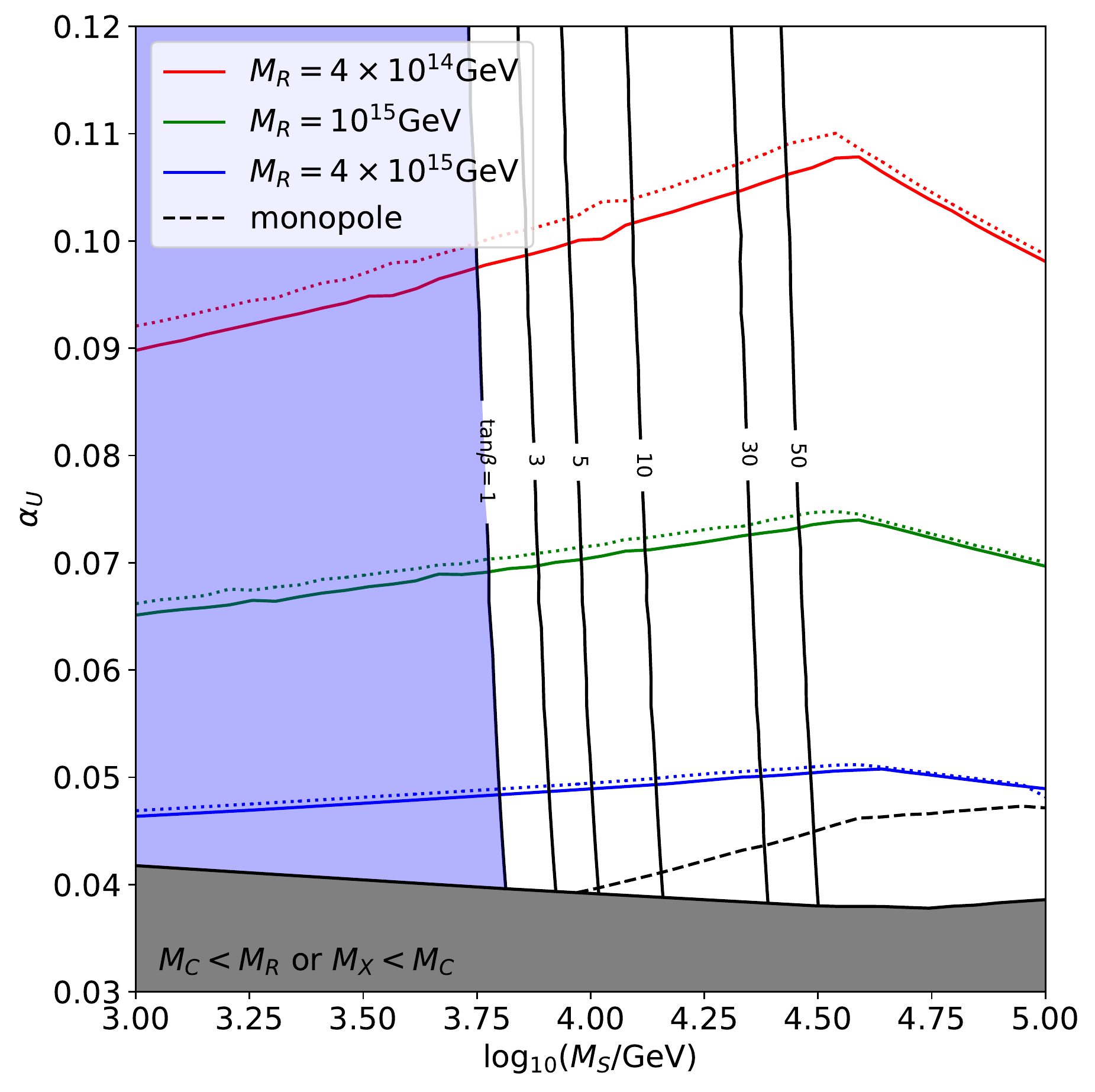}
\end{minipage}
\begin{minipage}[t]{0.32\hsize}
    \centering
    \includegraphics[width=1\linewidth]{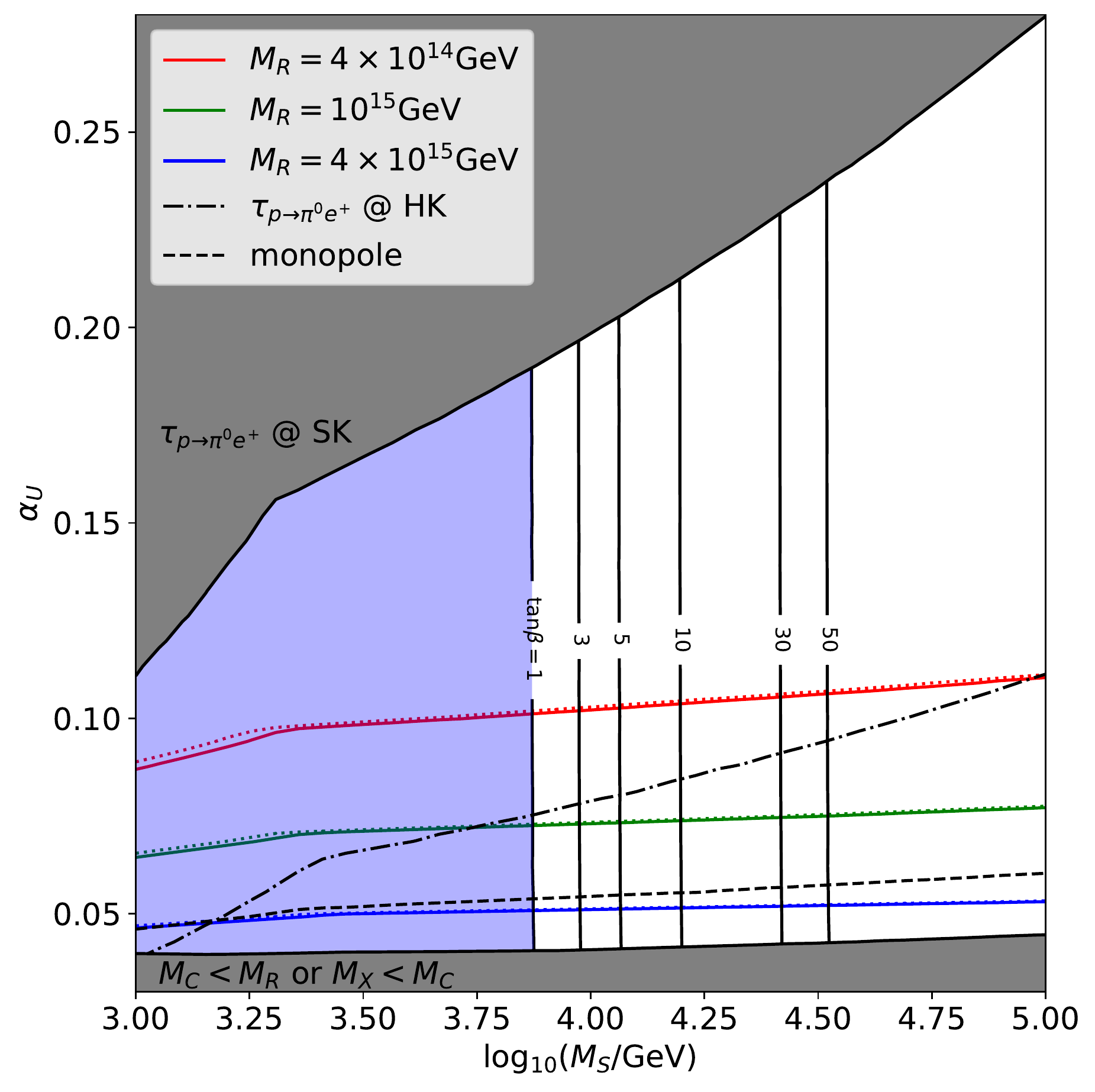}
\end{minipage}
    \caption{Contour plot of $M_R$ as a function of $M_S$ and $\alpha_U$.     
    {\em Left panel:} universal soft masses without threshold correction. {\em Middle panel:} universal soft masses with typical threshold corrections at the GUT-scale. {\em Right panel:} split SUSY spectrum with masses of gauginos fixed at $\mathcal{O}(1)\,\mathrm{TeV}$ and without threshold corrections.
    The red, green, and blue lines correspond to $M_R = 4\times 10^{14}\,\mathrm{GeV}$, $10^{15}\,\mathrm{GeV}$, $4\times 10^{15}\,\mathrm{GeV}$, respectively, with $\tan\beta=2$ (solid) and $50$ (dotted).
    The lower gray region is excluded for either $M_C < M_R$\,(abundant monopole) or $M_X < M_C$\,(unachievable unification).
    The upper gray and the left blue regions are excluded by the current lower bound on the proton partial lifetime $\tau_{p\to\pi^{0}e^{+}}$ and $\tau_{p\to K^{+}\bar{\nu}}$, respectively.
    The regions to the left of vertical lines are excluded by $\tau_{p\to K^{+}\bar{\nu}}$ depending on the choice of $\tan\beta=1,3,5,10,30,50$ from left to right.
    The dot-dashed line represents the future sensitivity on $\tau_{p\to\pi^{0}e^{+}}$ at the Hyper Kamiokande.
    The region below the dashed line leads to excessive monopole abundance without the realization of a supercooling phase in the $G_{3221}\to G_{\mathrm{SM}}\times M$ phase transition.
    \label{fig:modelA}
}
\end{figure*}

Fig.~\ref{fig:modelA} shows contours of the intermediate scale $M_R$ as a function of $M_S$ and $\alpha_U$.
The results in the left and the middle panels are obtained with universal soft masses, while the result in the right panel is obtained for a split spectrum with gaugino masses fixed at $\mathcal{O}(1)\,\mathrm{TeV}$ with other superpartners of SM particles at $M_S$.
In the middle panel, we take account of typical sizes of GUT-scale threshold corrections (see appendix for the details).
The colored contours correspond to the choices $M_R = 4\times 10^{14}\,\mathrm{GeV}$ (red), $10^{15}\,\mathrm{GeV}$ (green), and $4\times 10^{15}\,\mathrm{GeV}$ (blue), while the solid and dotted lines correspond to $\tan\beta=2$ and $50$, respectively.
The dependence on $\tan\beta$ comes from the two-loop contribution to RG equations from Yukawa couplings. Note that in a concrete model, $\tan\beta$ is constrained by the fermion mass spectrum.  Nevertheless, it has little impact on RG running and we will use $\tan\beta=2$ as a representative value.
The lower gray region is excluded by either $M_C < M_R$ or $M_X < M_C$. In the former case, monopoles are produced alongside cosmic strings. In the latter case, unification cannot be achieved.
All the viable parameter space in the figure leads to the unification scale $M_X$ of
$\mathcal{O}(10^{16}) \,\mathrm{GeV}$ or above.

{\bf Proton Decay.--}
As is often the case for SUSY GUT models,
the proton lifetime imposes severe constraints on the model parameter space.
The proton decay can be induced by the dimension-5 operators from the colored Higgs exchange\,\cite{Sakai:1981pk,Weinberg:1981wj,Dimopoulos:1981dw}.
Contrary to SUSY $SU(5)$ models, the existence of several colored Higgs multiplets in the present model
leads to various decay branches such as $p\to K^0 \ell^{+}$ as well as the popular $p\to K^{+} \bar{\nu}$ mode \cite{Babu:1998wi,Goh:2003nv}.
However, since the current constraints on such unusual decay modes are generally weaker than that on the $p\to K^{+} \bar{\nu}$ mode, we focus on the latter.
The proton lifetime is roughly given by \cite{Hisano:2013exa}
\begin{align}
    \tau_{p\to K^{+} \bar{\nu}} \sim 10^{35}\,\mathrm{yrs} \times 
    \sin^4 2\beta
    \left( \frac{M_S}{10^5\,\mathrm{GeV}} \right)^2
    \left( \frac{M_X}{10^{16}\,\mathrm{GeV}} \right)^2,
    \label{eq:dim5-decay}
\end{align}
where we have equated the colored Higgs mass $M_{H_C}$ with $M_X$ for simplicity.

The dimension-5 proton decay rate is model dependent and Eq.~\eqref{eq:dim5-decay} should be regarded as a rough estimate with large theoretical uncertainty.
Firstly, the size of the coupling between a colored Higgs and SM fermions is determined by the Yukawa coupling at the GUT scale, which is a source of the $\tan\beta$ dependence.
However, values of Yukawa couplings highly depend on the structure of the Yukawa sector in terms of $SO(10)$ superfields, while we use typical values of Yukawa couplings obtained by running the MSSM RG equations in the estimation.
Secondly, the intermediate scales alter the running of Wilson coefficients from those in SUSY $SU(5)$ models\cite{Hisano:2013exa} with which we perform our calculation.
However, this effect is not significant due to the proximity between $M_R$ and $M_X$, and is overwhelmed by the uncertainty in the mass spectrum of sfermions.
Furthermore, Eq.~\eqref{eq:dim5-decay} depends on the masses of the colored Higgses in $\bf 10$ and $\bf 120$ that can differ from the scale $M_X$ by an $O(1)$ factor.
As a whole, we expect $\mathcal{O}(1)$ uncertainties on our calculation of the proton decay rate
induced by colored Higgses.

Another important contribution to proton decay comes from dimension-6 operators induced by the heavy gauge boson exchange\,\cite{Weinberg:1979sa,Langacker:1980js}.
Relevant gauge bosons are those that transform under $G_{\mathrm{SM}}$ as $(3,2)_{5/6}$ and $(3,2)_{-1/6}$.
Compared with SUSY SU(5) models, the number of gauge bosons that contribute to the proton decay doubles, and the proton decay width to the most important decay mode, $p\to \pi^0 e^{+}$, increases.
The proton lifetime is given by \cite{Nath:2006ut}
\begin{align}
    \tau_{p\to \pi^0 e^{+}} \sim 5\times 10^{34}\,\mathrm{yrs} \times
    \left( \frac{0.04}{\alpha_U} \right)^2
    \left( \frac{M_X}{10^{16}\,\mathrm{GeV}} \right)^4,
\end{align}
where we equated the heavy gauge boson masses with $M_X$ for simplicity.
In the evaluation, we have used the RGE factor of Wilson coefficients valid for the MSSM running up to $M_X$ \cite{Hisano:2013ege}, which can be slightly modified due to the difference of RGEs above $M_R$.

In Fig.~\ref{fig:modelA}, we show the current lower limit on the proton lifetime
from the Super-Kamiokande experiment, $\tau_{p\to K^{+} \bar{\nu}} > 5.9\times 10^{33}\,\mathrm{yrs}$ \cite{Abe:2014mwa}, with black vertical lines.
The lines correspond to $\tan\beta=1,3,5,10,30$ and $50$ from left to right.
The region to the left of each line with smaller $M_S$ is excluded for each $\tan\beta$.
Therefore, the left blue region, whose right boundary corresponds to $\tan\beta=1$, is excluded for all $\tan\beta\geq 1$.
The upper gray region is the parameter space that is excluded
by the current limit of $\tau_{p\to \pi^0 e^{+}} > 1.6\times 10^{34}\,\mathrm{yrs}$~\cite{Miura:2016krn},
while the dot-dashed line represents the prospect of Hyper Kamiokande~\cite{Abe:2018uyc}.
Note that the GUT scale is generally large in the middle panel of Fig.\ref{fig:modelA}. As a result, the constraint from $p\to \pi^0 e^{+}$ is absent in this panel.

{\bf Monopole density.--}
In a series of phase transitions, $SO(10)\to G_{422}\to G_{3221} \to G_{\mathrm{SM}}\times M$,
we assume that inflation occurs before the $G_{3221} \to G_{\mathrm{SM}}\times M$ phase transition
and the reheating temperature $T_{R}$ is above the critical temperature of this phase transition
so that cosmic strings that emit GW are populated in the Universe.
On the other hand, the breaking of $G_{422}\to G_{3221}$ generates intermediate scale monopoles with
mass $m_{m}\simeq M_C/\alpha_4$\,\cite{Lazarides:1980va,Preskill:1984gd}.
To avoid the overclosure of the Universe, we require $T_{R}<M_C$ so that
the $G_{422}$ symmetry is not restored during the reheating and monopoles
are not generated by the Kibble mechanism\,\cite{Kibble:1976sj}.
However, when $T_R$ is not much lower than $M_C$, monopoles may still be produced through annihilation of particles in the thermal bath
with a suppressed rate\,\cite{Preskill:1979zi,Turner:1982kh}.
For $m_{m}/T_{R}\lesssim20$, the monopole is thermalized
and its relic density overcloses the Universe.
For $m_{m}/T_{R}\gtrsim20$, the monopole
is produced in out-of-equilibrium with a final density,
\begin{equation}
\frac{n_{m}}{n_{\gamma}}\simeq3\times10^{3}
\left(\frac{m_{m}}{T_{R}} \right)^3 e^{-\frac{2m_{m}}{T_{R}}}\,,
\end{equation}
where $n_\gamma$ is the photon number density.
Such monopole density is constrained in several ways. The dark matter
cannot be solely made of monopoles because the local monopole density
would be in severe contradiction with the null result of monopole
searches such as MACRO\,\cite{Ambrosio:2002qq}. We can then require $\Omega_{m}\ll\Omega_{M}$,
where $\Omega_{m}$ and $\Omega_{M}$ are the monopole and dark
matter densities normalized by the critical density of the Universe.
This sets $m_{m}/T_{R}\gtrsim40$ for $m_{m}\sim10^{16} \, \rm GeV$.
Another equally strong constraint comes solely from the direct search of monopoles by MACRO.
Assuming monopoles are accelerated fast enough by the galactic magnetic field\,\cite{Parker:1970xv},
they are not bounded to galaxies and are uniformly distributed in the Universe, with
a flux near the Earth,
\begin{equation}
F_{m}\simeq1.4\times10^{-14}{\rm cm}^{-2}{\rm s}^{-1}\Omega_{m}\left(\frac{10^{16}\,{\rm GeV}}{m_{m}}\right)\left(\frac{v_{m}}{300\,{\rm km\,s^{-1}}}\right),
\end{equation}
where $v_{m}$ is the relative velocity between the monopole and the Earth.
The MACRO monopole search puts a constraint on the flux,
$F_{m}\lesssim1.8\times10^{-15}{\rm cm}^{-2}{\rm s}^{-1}$,
which again sets the limit $m_{m}/T_{R}\gtrsim40$. Assuming the critical temperature $T_c$
of the $G_{3221} \to G_{\mathrm{SM}}\times M$ phase transition is around the symmetry breaking scale,
$T_R\gtrsim T_{c}\sim M_R$ is required for the production of cosmic strings. We thus find a constraint on the hierarchy between the two intermediate scales,
$M_C/T_R\gtrsim M_C/M_R \gtrsim 40 \alpha_4$.
This constraint is shown in Fig.~\ref{fig:modelA} with dashed lines; the regions below the dashed lines are constrained.

If the $G_{3221} \to G_{\mathrm{SM}}\times M$ phase transition is of the strong first-order
and experiences a supercooling phase,
produced monopoles are diluted and the monopole density is suppressed.
In this case, the constraint discussed above is irrelevant.
The abundance of cosmic strings produced at such a first order phase transition may be different from
that of a second-order transition.
However, it hardly affects the GW generation because it has been known that
a network of cosmic strings reaches a scaling solution 
\cite{Kibble:1984hp}.

{\bf Discussions.--}
For a given mass scale of supersymmetric particles $M_S$, a precise gauge coupling unification relates
the intermediate scale $M_R$ of cosmic strings to the $SO(10)$ breaking scale $M_X$ or the size of the unified gauge coupling $\alpha_U$.
The high cosmic string scale $M_R\gtrsim 10^{14}$\,GeV inferred by the NANOGrav signal is a generic feature of the SUSY SO(10) model as we demonstrate in Fig.~\ref{fig:modelA}.
The left panel of Fig.~\ref{fig:modelA}, which corresponds to the case with universal soft masses, shows that the current constraint on the proton lifetime and also the monopole overproduction constraint require
supersymmetric particles at a low-energy scale
under the condition of the precise gauge coupling unification without the GUT-scale threshold corrections.
There is a tiny allowed parameter region with $M_S = \mathcal{O}(10)\,\mathrm{TeV}$
and $M_R = \mathcal{O}(10^{15})\,\mathrm{GeV}$ which is consistent with the NANOGrav data.
However, this region with $\tan \beta \lesssim 3$ is in tension with the observed SM Higgs mass
\cite{Bagnaschi:2014rsa}.
As suggested in the middle panel, the GUT-scale threshold corrections can change this conclusion completely.
The GUT scale $M_X$ tends to be higher, which renders the $p\to\pi^0 e^{+}$ decay unobservable unless $\alpha_U = \mathcal{O}(1)$.
There is a vast parameter region unconstrained by current and future proton decay experiments.
In this case, the GW observation together with a better understanding of cosmic string formation serves as the only means to narrow down the scale $M_R$ and thus the $SO(10)$ breaking scale $M_X$ and the size of the unified gauge coupling $\alpha_U$ through the requirement of gauge coupling unification.
For the case of split SUSY without threshold correction, the right panel of Fig.~\ref{fig:modelA} also shows a large allowed region with $M_S \gtrsim 10\,\mathrm{TeV}$,
which can be consistent with the SM Higgs mass while explaining the stochastic GW signal.
The search for the proton decay at the Hyper-Kamiokande will explore a large fraction of the allowed parameter space, and the decay will be observed when $M_R = \mathcal{O}(10^{14})\,\mathrm{GeV}$, which is favored by the recent numerical simulation of cosmic strings to explain the NANOGrav data\,\cite{Ellis:2020ena}, with a moderate choice of $M_S$.
Because of the high cosmic string scale, the monopole abundance also places non-trivial constraints for the low $\alpha_U$ region, even if its initial density is diluted by inflation. This complements the proton decay constraints at the high $\alpha_U$ region.
In conclusion, the GW observation gives us a way to probe the supersymmetric grand unification.
To extract the intermediate scale $M_R$ from the GW signal precisely,
it is essential to reduce the uncertainty for the initial loop size $\alpha$ in the cosmic string spectrum.


\section*{Acknowledgements}
S.C. is supported by JSPS KAKENHI grant.
Y.N. would like to thank Kohei Fujikura, Motoo Suzuki and Tsutomu T. Yanagida for discussions.
Y.N. is grateful to Kavli IPMU for their hospitality during the COVID-19 pandemic.
J.Z. is supported in part by the National NSF of China under grants 11675086 and 11835005.

\vspace{0.5cm}

{\bf Appendix: one intermediate scale.--}
The simplest possibility of the multi-step $SO(10)$ breaking is the case with one intermediate scale,
$SO(10) \to H \to G_{\mathrm{SM}} \times M$.
We focus on models with $H=G_{3221}$ or $G_{421}$
where cosmic strings are formed at the intermediate scale $M_R$ as shown in Table~\ref{tab:vacuum}.

The hierarchical VEVs, $|s|\sim |a|\sim M_X \gg |\sigma| \sim M_R \gg |b| \sim M_2 = M_R^2/M_X$,
lead to the breaking pattern of $SO(10) \to G_{3221} \to G_{\mathrm{SM}} \times M$.
The mass spectrum of this model is obtained by taking the limit of $M_C \rightarrow M_X$
in Table~\ref{tab:mass-spectrum}.
As in the case of two intermediate scales,
the gauge couplings of $SU(2)_R$ and $U(1)_{B-L}$ are matched to that of the SM hypercharge
at $M_R$ through Eq.~\eqref{eq:matching-3221}.
All the gauge couplings of $G_{3221}$ run further to the unification scale $M_X$
and unify into a unique value $\alpha_U$.
Consequently, we obtain the one-loop level relationship between
the gauge coupling constants $\alpha_i$ at $M_Z$ and $\alpha_U$ as
\begin{align}
    \frac{2\pi}{\alpha_i (M_Z)} &= \frac{2\pi}{\alpha_U} + \left[
    b_i^{(1)} \ln\frac{M_S}{M_Z}
    + b_i^{(2)} \ln\frac{M_2}{M_S} \right. \notag \\
    &\left.
    + b_i^{(a)} \ln\frac{M_R}{M_2}
    + b_i^{(b)} \ln\frac{M_X}{M_R}
    \right],
    \label{eq:gcu-2}
\end{align}
where $b_i^{(1)}$ and $b_i^{(2)}$ are given in Table.~\ref{tab:RGE}, while $b_i^{(a)}=(57/5,1,-3)$ and $b_i^{(b)}=(9,1,-3)$.
We solve the three equations \eqref{eq:gcu-2} in terms of the three parameters $M_R$, $M_X$, and $\alpha_U$, and obtain a set of solutions as a function of $M_S$.
It is found that the correct hierarchy $M_Z < M_S < M_R < M_X$ requires
$M_S \lesssim 1\,\mathrm{TeV}$, which is already excluded by collider searches.

The breaking pattern of $SO(10) \to G_{421} \to G_{\mathrm{SM}} \times M$
is realized by the hierarchical VEVs, $|s| \sim |b| \sim M_X \gg |\sigma| \sim M_R \gg |a| \sim M_2$.
In this setup, there are light degrees of freedom described as $(6,1)_{\pm 4/3}$ and $(1,1)_{\pm 0}$ under $G_{\mathrm{SM}}$, all of which have masses of $\mathcal{O}(M_2)$, in addition to the to-be Nambu-Goldstone bosons with masses of $\mathcal{O}(M_R)$.
At the intermediate scale $M_R$, the matching between the gauge coupling constants is given by
\begin{align}
    &\alpha_4 (M_R) = \alpha_3 (M_R), \\
    &\frac{2}{5}\alpha_4^{-1} (M_R) + \frac{3}{5}\alpha_{1R}^{-1} (M_R) =  \alpha_1^{-1} (M_R),
\end{align}
where $\alpha_{1R}$ denotes the $U(1)_R$ gauge coupling.
The RG evolution of the gauge couplings is again governed by Eq.~\eqref{eq:gcu-2}, though in this case $b_i^{(a)}=(97/5,1,2)$ and $b_i^{(b)}=(81/5,1,0)$.
Again we found that the hierarchy $M_Z < M_S < M_R < M_X$ requires $M_S < 1\,\mathrm{TeV}$, which is already excluded.

{\bf Appendix: threshold corrections.--}
We here estimate threshold corrections to the couplings at the GUT scale $M_X$
from the spectrum of $S$, $A$, $\Sigma$ and $\bar\Sigma$.
If we ignore $\bf 10$ and $\bf 120$, the theory is defined by the terms in Eq.~\eqref{eq:SO10_lagrangian}. Applying the vacuum conditions, the mass parameters $m_S$, $m_A$, and $m_\Sigma$ can be traded with the VEVs $s$, $a$ and $\sigma$. The remaining free parameters are the couplings $\lambda$, $\lambda_S$, $\eta_S$, $\bar\eta_S$ and $\eta_A$. The threshold corrections can then be parametrized by these dimensionless couplings and the VEVs. The one-loop contribution to the running coupling $\alpha_i^{-1}(Q)$ at $Q\gtrsim M_X$ from all chiral superfields with the same $G_{\rm SM}$ representation $R$ is $2\pi\Delta\alpha_i^{-1}(Q) = \sum_{j} b_i^R \ln\frac{m_j}{Q}$. 
$b_i^R = l_i^R$ is the Dynkin index of the representation $R$. Since these superfields have the same SM quantum number and mix with each other, their mass terms are generally described by a non-diagonal mass matrix $M(R)$ after intermediate symmetry breakings, as given in the appendix of \cite{Melfo:2010gf}. Neglecting Nambu-Goldstone bosons\footnote{The symmetry breaking scale
is defined as the mass of gauge bosons so Nambu-Goldstone bosons do not contribute to the threshold correction.},
the contribution can be evaluated as
\begin{equation}
2\pi\Delta\alpha_i^{-1}(Q)
=
b_i^R \ln \left|\frac{a_k(M(R))}{Q^{n-k}} \right|\,,
\end{equation}
where $n$ is the dimension of the mass matrix $M(R)$, $k$ is the number of zero eigenvalues that correspond to the Nambu-Goldstone bosons, and $a_k(M(R))$ is the coefficient of the $x^k$ term of the characteristic polynomial $|{\rm Det}(M(R)-x{\bf 1})|$. For comparison, the 1-loop step-wise contribution to the running coupling is 
$2\pi\Delta_0 {\alpha}_i^{-1}(Q) = \sum_{j} b_i^R \ln\frac{\tilde Q_j}{Q}$, where $\tilde Q_j\in \{M_1,\,M_2,\,M_R,\,M_C,\,M_X\}$ is the mass scale of the particle $j$. The threshold correction $\lambda_i\equiv 2\pi\sum_R\left(\Delta\alpha_i^{-1}(Q) - \Delta_0\alpha_i^{-1}(Q)\right)$  is then,
\begin{equation}
\lambda_i
=
\sum_R
b_i^R \ln \left|\frac{a_k(M(R))}{\Pi_{j\neq {\rm NG}}\, \tilde Q_{j} }  \right|\,,	
\label{eq:thres_corr}
\end{equation} 
with NG stands for Nambu-Goldstone bosons. For gauge coupling unification, it is more convenient to calculate
\begin{equation}
\Delta\lambda_{ij}\equiv \lambda_i - \lambda_j\,.
\end{equation}
Given the mass matrices in \cite{Melfo:2010gf}, the calculation $\Delta\lambda_{ij}$ is straightforward with Eq.\eqref{eq:thres_corr}.
For our model A, identifying $M_X=s$, $M_C=a$, $M_R=\sigma$, and in the limit of $s\gg a \gg \sigma$, we obtain
\begin{equation}
\Delta\lambda_{12}
\simeq -7.4
-\frac{2}{5}\left(
16\ln \lambda -5\ln\eta_A -2\ln\xi^3_S
\right),
\end{equation}
and
\begin{align}
\Delta\lambda_{13}
\simeq 
\begin{cases}
-9.0
-\frac{3}{5}\left(
4\ln \lambda -5\ln\eta_A +2\ln\xi^3_S
\right), \\[1ex]
\qquad \qquad \qquad \qquad \qquad \text{for $\sigma^2/a > a^2/s$,}
 \\[1ex]
-9.0
+\ln\frac{a^3}{s\sigma^2}
-\frac{3}{5}\left(
4\ln \lambda -5\ln\eta_A +2\ln\xi^3_S
\right), \\[1ex]
\qquad \qquad \qquad \qquad \qquad \text{for $\sigma^2/a < a^2/s$,}
\end{cases}
\end{align}
where we have defined
$\xi^3_S\equiv\eta_S \bar\eta_S \lambda_S$.
In the middle panel of Fig.~\ref{fig:modelA},
we define the gauge coupling at the GUT scale as
\begin{align}
    \alpha_1 (M_X) &= \alpha_U,\\
    \alpha_i (M_X) &= \alpha_U \left( 1+\frac{\alpha_U}{2\pi} \Delta \lambda_{1i} \right)~~(i=2,3),
\end{align}
with $\Delta\lambda_{12} = -7.5$ and $\Delta\lambda_{13}=-9.0$, which are typical values observed when all the dimensionless couplings are equal to unity.

\bibliography{bib}
\bibliographystyle{utphys}

\end{document}